\documentclass[english,twocolumn]{article}

\usepackage[big]{dgruyter}

\usepackage{amsmath, amssymb}
\usepackage{cases}
\usepackage{subfig}
\usepackage{caption}
\newcommand{\Webe}{\operatorname{\mathit{W\kern-.24em e}}}

\begin{document}

  \articletype{Research Article{\hfill}Open Access}

  \author*[1]{Akio Nishimura}

  \author[2]{Henry Weller}
  \author[3]{Hirokazu Maruoka}
  \author[3]{Masao Takayanagi}
  \author[3]{Hideharu Ushiki}

  \affil[1]{United Graduate School of Agricultural Science, Tokyo University of Agriculture and Technology, 3-5-8 Saiwaicho, Fuchu, Tokyo 183-8509, Japan, E-mail: akio@akionux.net}
  \affil[2]{CFD Direct Ltd., United Kingdom}
  \affil[3]{United Graduate School of Agricultural Science, Tokyo University of Agriculture and Technology, 3-5-8 Saiwaicho, Fuchu, Tokyo 183-8509, Japan}

  \title{\huge Dynamics of a dry-rebounding drop: observations, simulations, and modeling}

  \runningtitle{Dynamics of a dry-rebounding drop}
  \runningauthor{A. Nishimura \textit{et al.}}


  \begin{abstract}
      {Dynamics of a dry-rebounding drop was studied experimentally, numerically, and theoretically.
      Experimental results were reproduced by our computational fluid dynamics simulations, from which time series of kinetic energy, potential energy, and surface energy were obtained.
      The time series of these energies quantitatively clarified the energy conversion and loss during the dry-rebound.
      These results were interpreted by using an imaginary spring model and a spherical harmonic analysis.
	  The spring model explained the vertical deformation of the drop, however, could not completely explain the energy loss; the timings of the energy loss did not match.
	  From a viewpoint of the spherical harmonic deformation of a drop, the deformation of the drop after the impact was found to be a combination of two vibrational motions. One of the two vibrational motions is an inertial motion derived from the free-fall and the another is a pressure-induced motion derived from a pressure surge due to the sudden stop of the bottom part of the drop at the impact.
	  The existence of the pressure surge at the impact was confirmed in the simulated results.
	  The pressure-induced motion resists the inertial motion and consequently dumps the kinetic energy of the drop.
	  }
\end{abstract}

  \journalname{Open Phys.}

\DOI{10.1515/phys-2018-0039}
  \startpage{1}
  \received{Nov 10, 2017}
  \revised{}
  \accepted{Mar 01, 2018}

  \journalyear{2018}
  \journalvolume{}
  \journalissue{}

\maketitle
\section{Introduction}
A drop on a superheated surface, of which the temperature is higher than a critical point, floats on a stable vapor film generated by evaporation of the drop.
This is referred to as the Leidenfrost effect, named after the person who first discovered it \cite{leidenfrost1966,quere2013}.
We will thus refer to such a drop as a Leidenfrost drop.

The drop can be regarded as completely floating and non-wetting on the surface \cite{biance2003} and the heat flux from the surface towards the drop is sufficiently small that changes in the fluid properties are negligible due to the heat insulating effect of the vapor \cite{biance2006,bertola2014}.
If the drop falls from a certain height towards the heated surface, it bounces on the surface similar to a bouncing ball, which is referred to as a dry rebound \cite{bertola2014}.
In a dry rebound, the drop falls while converting the initial potential energy to kinetic energy,
then impacts on the surface while converting the kinetic energy to surface energy by deformation to a disk-like shape,
and then shrinks while converting the surface energy to kinetic energy again.
In these dynamics, the drop behaves similar to an elastic spring.
As such, the spring model has helped to reveal interesting characteristics of the drop \cite{okumura2003,biance2006}.

A small amount (in the order of 100 ppm) of polymer additives in the drop is known to change the dynamic behavior and energy loss of the drop during the bounce \cite{bertola2004,bertola2009,smithBertola2010,bertola2010,bertola2014}.
However, the energy conversion and loss of a dry-rebounding drop, even without the polymer additive, have remained unclear \cite{biance2006}.

An efficient approach to understand the drop characteristics is a numerical simulation of the drop under a completely non-wetting condition, which has successfully reproduced the experimental drop results \cite{renardy2003}.
Computational fluid dynamics (CFD) not only enables unknown phenomena to be expected, but also facilitates clarification of detailed physical information regarding complex fluidic phenomena.
Previous studies \cite{naoe2014,roohi2013,berberovic2009} have shown that the volume of fluid (VOF) method provides reliable and reasonable results on two-phase flows.
For example, an impacting mercury drop \cite{naoe2014}, a cavitation around a two-dimensional hydrofoil \cite{roohi2013}, and a drop impacting onto a liquid layer of finite thickness \cite{berberovic2009} have been simulated with the VOF method.

In the present study, dynamics of a dry-rebounding drop is observed by a high-speed camera, numerically simulated with a CFD solver, and theoretically modeled with a damped spring model.
We focus on the first few bounces whose time span ($\sim 0.1$ s) is much shorter than the life time of a Leidenfrost drop on a hot plate at a temperature of 380 ${}^\circ$C ($\sim 100$ s) \cite{biance2003}, thus the volume change due to the evaporation is negligible.
In the experiments (Section 2), drops falling from different heights were captured with a high-speed camera and the videos were analyzed to measure the geometrical properties of the drops.
CFD simulations of the dry-rebounding drops were then performed using a two-phase solver under a completely non-wetting condition on a flat plate (Section 3).
In Section 4, the numerical results were assessed by comparison with the experimental result, and then time evolutions of the kinetic energy, potential energy, and surface energy of the drop were calculated.
An imaginary damped spring model and a spherical harmonic analysis were introduced to elucidate the mechanism for the energy loss of the drop.
Final conclusions are described in Section 5.

\section{Experiment}
\begin{figure}[h]
    \begin{center}
        \includegraphics[width=7.4cm]{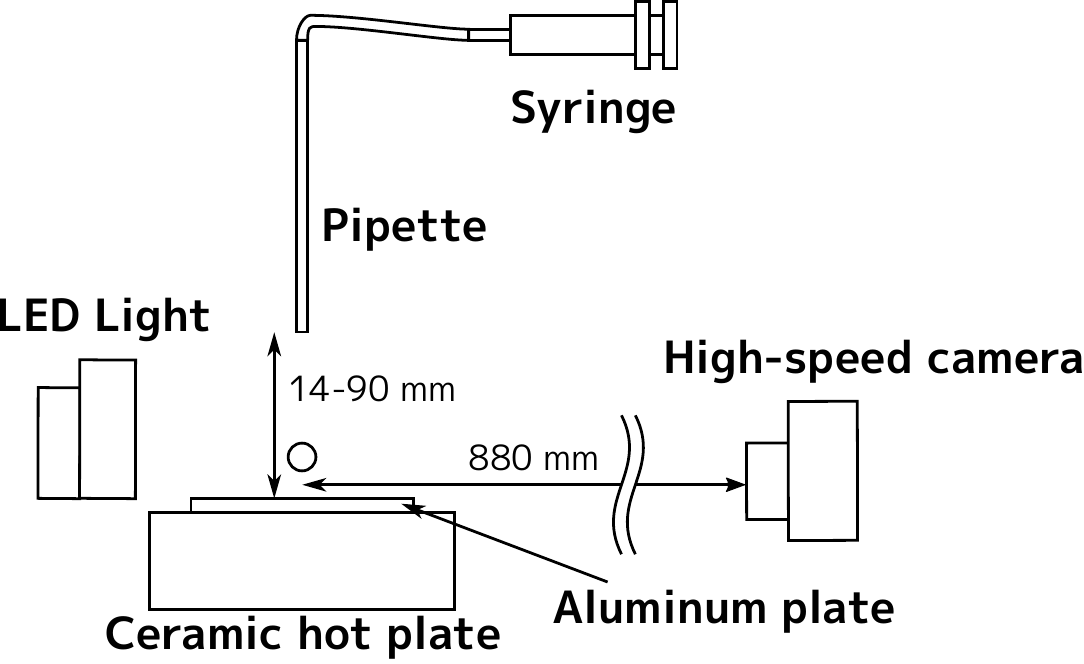}
        \caption{Experimental setup used to capture drop impact on a superheated flat surface. The drop was generated at the tip of a pipette and captured with a high-speed camera.}
        \label{fig:expSetup}
    \end{center}
\end{figure}

The experimental setup is shown in Figure\ \ref{fig:expSetup}.
An aluminum plate ($100\; \mathrm{mm} \times 100\; \mathrm{mm} \times 5\; \mathrm{mm}$)  was heated on a ceramic hot plate (As one, CHP-170DN) to 400 $^{\circ}\mathrm{C}$, which is sufficiently higher than the Leidenfrost temperature and the boiling point of water.
A drop of distilled water was dropped from a pipette (inner diameter: $1 \mathrm{mm}$).
Averaged diameter of the drop in all the experiments was 3.69 mm with the standard error of 0.10 mm.
Experiments were conducted under the room conditions; the temperature was $23.9\pm0.3\;^{\circ}$C and the humidity was $80\pm4\;$\%RH.

To characterize the drop, the Weber number $\Webe$ was used, which
is a non-dimensional number that gives the ratio of kinetic energy to the surface energy of the drop and thus represents the stability of a drop.
Weber number especially for a drop at impact is referred to as the dynamic Weber number, and can be expressed as
\begin{equation}
	\Webe = \frac{\rho U_{\mathrm{impact}}^2 D_0}{\sigma},
	\label{eq:weber number}
\end{equation}
where $\rho$ is the density of the drop, $U_{\mathrm{impact}}$ is the velocity of the drop just before impact, $D_0$ is the initial diameter of the drop, and $\sigma$ is the surface tension coefficient between air and the internal fluid of the drop.

The height of the pipette tip above the plate was adjusted from 9 mm to 25 mm at 1 mm increments to change the impact velocity of the drop, $U_\mathrm{impact}$.

The drop impact on the plate was captured using a high-speed camera (Casio, EXILIM EX-F1) with a frame rate of 1200 fps and a resolution of 336$\times$96 pixels.
Experiments were performed three times for each initial height.

\begin{figure}[h]
    \begin{center}
        \includegraphics[width=8.0cm]{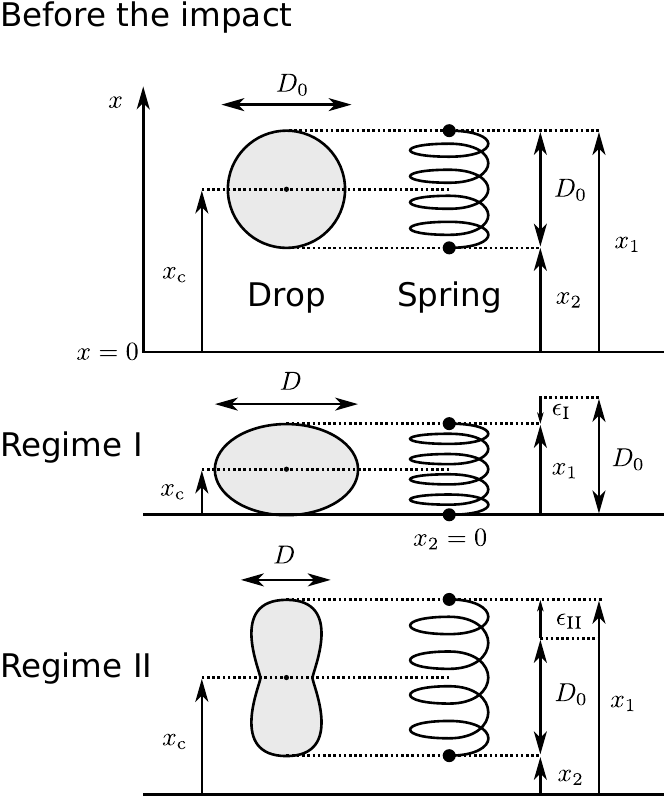}
		\caption{Schematic diagram of a drop and the imaginary damped spring model prior to impact, in regime I, and in regime II. The geometric properties of the drops were measured using an image-processing pipeline. The spring has two mass points (each one is half the weight of the drop) at both ends.}
        \label{fig:dropGeom}
    \end{center}
\end{figure}

\begin{figure}[h]
	\begin{center}
        \includegraphics[width=3.0in]{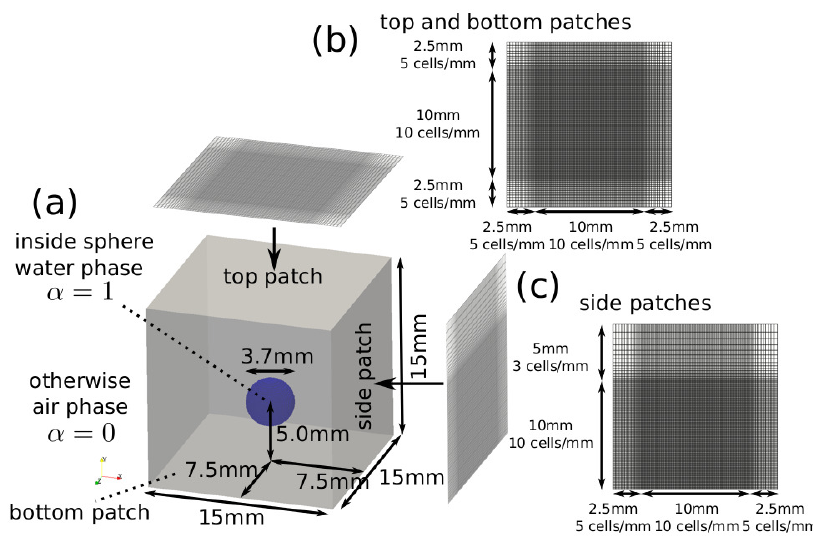}
        \caption{Schematic diagram of the CFD setup. (a) System size and initial setup of the volume fraction of water, $\alpha$. The grid mesh from (b) top and bottom views, and (c) side views. The central cubic domain has the finest and unity resolution, otherwise domain has reduced resolution with expanding cell size outward.}
		\label{fig:cubicMesh}
	\end{center}
\end{figure}

\begin{table}[h]
	\caption{Boundary conditions used in the calculation.
    "Bottom patch" is the patch on which the drop impacts and "Other patches" are placed at the left, the right, the back, the front, and the upper sides of the drop.
		Under condition \#1, the gradient value of the boundary field is fixed to zero, except on the tangential component which is set to 0 for inflow.
		Under condition \#2, the velocity field on the patch is evaluated from the flux, switching zero gradient, and the fixed value, depending on the direction of velocity with respect to the boundary.
	Under condition \#3, the pressure gradient was adjusted depending on the flux.}
	\label{tb:boundaryConditions}
	\begin{tabular}{ccc}
		\hline
		Variable & Bottom patch & Other patches \\
		\hline
		\rule{0pt}{2.5ex}$\mathbf{U}$ & \#1 & \#1 \\
		$\alpha$ & non-wetting condition & \#2 \\
		$p-\rho gh$ & \#3 & fixed value ($10\; \mathrm{kPa}$)\\
		\hline
	\end{tabular}
\end{table}

Captured videos were processed using an image processing pipeline that was written in Python \cite{python} and using OpenCV \cite{opencv}, an open source computer vision library.
The images were processed into binary images that indicate the interior or exterior of the drop, and from the binary images, contours for the drop edge were obtained.
The geometric properties of the drop were obtained from the image processing pipeline (Figure\ \ref{fig:dropGeom}): top height $x_1$, bottom height $x_2$, and width $D$.

\section{Numerical Method}
\subsection{Finite volume method and volume of fluid method}
The drop was simulated using a two-phase solver, interFoam \cite{interFoam} of OpenFOAM (version 3.0.x) \cite{of,of30x}, a CFD toolkit software that can be used and exploited under the GNU General Public License (GPL) \cite{gpl}.
The interFoam solver is based on the finite volume method (FVM).
In the FVM, the domain of calculation is divided into finite-volume cells which are referred to as control volumes, and physical values (e.g., velocity and pressure) are assigned to the centroid or faces of each cell.

A type of VOF method is used in interFoam to model two-phase flow and to track the free surface.
In the present simulation, the two phases of water and air were considered.
Note that in the two-phase flow, the volume fraction of liquid, $\alpha_\mathrm{l} = \alpha$, determines the volume fraction of gas, $\alpha_\mathrm{g} = 1-\alpha$.
\subsection{Interface capturing}
An efficient method is required to simulate a multiphase flow and capture a sharp interface between the two immiscible phases.
VOF methods have a problem with respect to the diffusive interface between two phases.
In VOF methods, the volume fraction of each phase is tracked through every control volume.
The volume fraction of each phase is expressed by a scalar function, which is referred to as a volume function or a color function.
To reproduce the interface between immiscible phases, the volume function needs to keep a steep gradient at the interface.
However, the steep gradient readily dissipates because VOF methods solve a momentum equation for a mixture of immiscible phases. 
Therefore, a special treatment is needed for the interface of volume functions.
The relative velocity, $\mathbf{U}_{r}$, is used to compress the interface between the two phases.
Weller \cite{weller2015} proposed a relative velocity between two phases, $\mathbf{U}_{r}$, as follows:
\begin{equation}
    \mathbf{U}_{r} = \mathrm{min}(C_{\alpha}|\mathbf{U}|, \mathrm{max}(|\mathbf{U}|)) \frac{\nabla \alpha}{|\nabla \alpha|},
\end{equation}
where $\mathbf{U}$ is the velocity field, and $C_\alpha$ is a coefficient set to 1 in the present simulation.
This method has proven to be reliable in maintaining a sharp interface \cite{weller2015}.

\subsection{Surface tension force}
Surface tension force is calculated using the continuum surface force (CSF) model \cite{brackbill1992}:
\begin{equation}
	\mathbf{F}_{\sigma} = \sigma \kappa \nabla \alpha,
\end{equation}
where $\sigma$ is the surface tension coefficient and $\kappa$ is the curvature of the interface between the liquid and gas.
$\kappa$ is given by
\begin{equation}
	\kappa = - \left( \nabla \cdot \hat{\mathbf{n}}\right),
	\label{eq:curvature}
\end{equation}
in which $\hat{\mathbf{n}}$ is the gradient vector at the face, which is given by
\begin{equation}
	\hat{\mathbf{n}} = \frac{\nabla \alpha}{|\nabla \alpha| + \delta_n},
\end{equation}
\begin{figure*}[ht]
    \includegraphics[width=\textwidth]{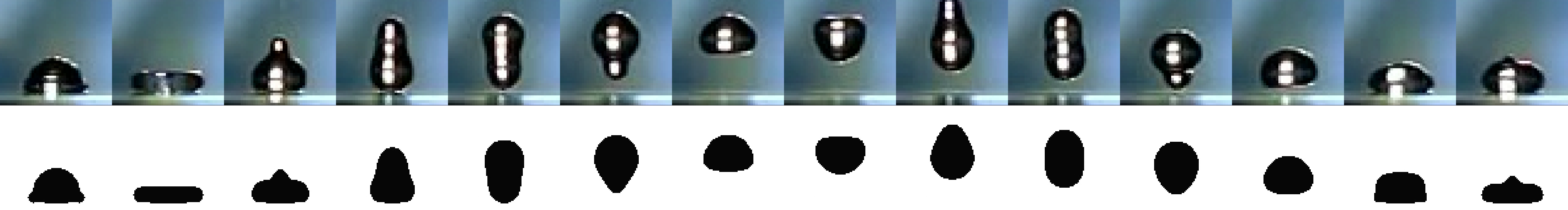}
    \includegraphics[width=\textwidth]{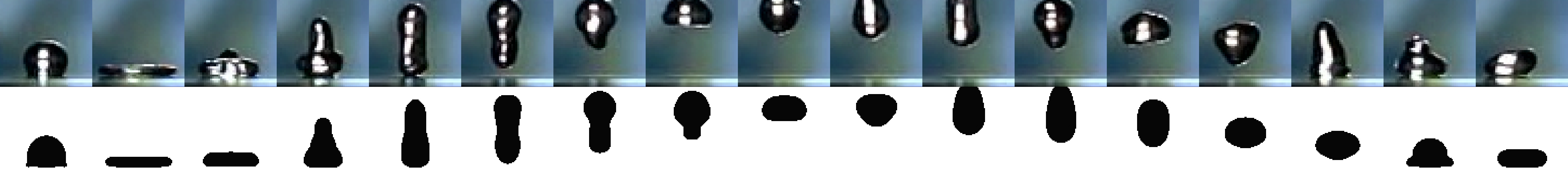}
    \includegraphics[width=\textwidth]{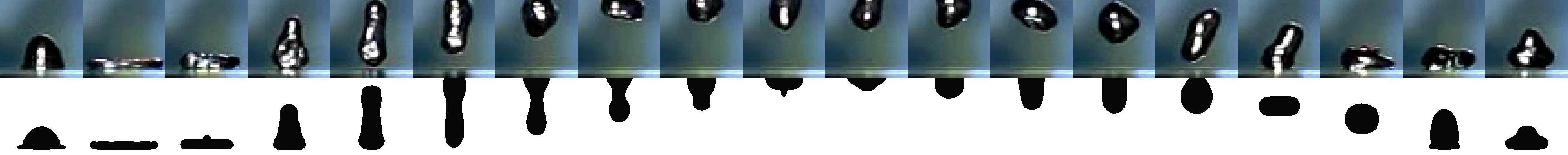}
    \caption{Sequential images of experimental drops (upper rows) and simulated drops (lower rows) captured every 5 $\mathrm{ms}$ for $\Webe = 7$ (top), $\Webe = 15$ (middle), and $\Webe = 23$ (bottom).}
    \label{fig:sequential images}
\end{figure*}
where $\delta_n$ is a stabilization factor depending on the volume of grid cells. The typical value of $\delta_n$ in our simulation is $1.0 \times 10^{-5} \mathrm{m}^{-1}$.

\subsection{Velocity-pressure coupling}
The momentum equation is given by
\begin{equation}
	\begin{split}
		\frac{\partial}{\partial t} (\rho \mathbf{U}) &+ \nabla \cdot (\rho \mathbf{U}\mathbf{U}) = \\
                                                &- \nabla p + \nabla \cdot \boldsymbol{\tau} + (\mathbf{g}\cdot \mathbf{h}) \nabla \rho + \mathbf{F}_{\sigma},
	\end{split}
	\label{eq:momentum equation}
\end{equation}
where $\rho$ is the mixture density, $p$ is the pressure, $\mathbf{g}$ is the gravity vector, $\mathbf{h}$ is the position vector in the vertical direction, $(\mathbf{g}\cdot \mathbf{h})\nabla \rho$ is the buoyancy force, and $\boldsymbol{\tau}$ is the deviatoric stress.
The interFoam solver uses the PIMPLE method, which is a combined velocity-pressure coupling algorithm of the SIMPLE (Semi-Implicit Method for Pressure-Linked Equations) and PISO (Pressure Implicit with Splitting of Operator) algorithm \cite{issa1986}.
The PIMPLE algorithm is summarized as the following routine.
\begin{enumerate}
	\item {\it Momentum prediction}: Predict the velocity field using the momentum equation.
	\item {\it Pressure solution}: Solve the pressure equation and correct flux.
	\item {\it Explicit velocity correction}: Correct the velocity field with the solved pressure field.
\end{enumerate}
The routine is repeated for certain number of times, which was two times in the present simulation.

\subsection{Computation and post-processing}
A diameter given by the average diameter of 51 experimental drops was adopted as the initial diameter of the numerical drop, $D_0 = 3.7\; \mathrm{mm}$.
The initial velocity of the drop was determined using the conservation of mechanical energy:
\begin{equation}
	U_\mathrm{impact} = \sqrt{2gx_\mathrm{c,0}},
\end{equation}
where $g$ is the gravitational acceleration, and $x_\mathrm{c,0}$ is the initial height of the centroid of the drop.
The viscosities of water and air were set to $1.0\times 10^{-3}\; \mathrm{Pa \cdot s}$ and $1.84\times 10^{-5}\; \mathrm{Pa \cdot s}$, respectively.
The field of the initial volume fraction of water was set to $\alpha=1.0$ at the interior of the drop and $\alpha=0.0$ at the outside of the drop.
The surface tension coefficient $\sigma$, between water and air was set to $0.07\; \mathrm{N \cdot m^{-1}}$.

The boundary conditions used for the calculation are shown in Table \ref{tb:boundaryConditions}.
The contact angle between water and air on the bottom patch on which the drop impacted was set to 180${}^{\circ}$ (a perfectly hydrophobic surface), which means that the gradient of $\alpha$ on the bottom boundary is determined as the negative normal vector of the boundary patch.
The schematic diagram of the CFD setup is shown in Figure\ \ref{fig:cubicMesh}.
The calculation domain was in the shape of a cube, 1.5 cm on a side (Figure\ \ref{fig:cubicMesh}(a)).
A spherical drop of 3.8 mm diameter was placed at a height of 5 mm above the bottom boundary (Figure\ \ref{fig:cubicMesh}(a)).
To improve efficiency of the calculation, the resolution of the mesh is uniform and finest at the interior of the central rectangular column covering the drop, while it becomes coarser toward the exterior, where the drop will never enter (Figures\ \ref{fig:cubicMesh}(b) and \ref{fig:cubicMesh}(c)).
Number of grid cells above 1.0 cm from the bottom patch was reduced to 30 \% of the finest region, and number of grid cells outside a square with a side length 1.0 cm and a center of the drop was reduced to 50 \% of the finest region.
From the finest grid cells towards boundaries except the bottom, the length of edge of grid cells are expanded with the expansion ratio of smallest length to largest length of the edge length of grid cells.
The expansion ratios are 2.81 towards the left, the right, the front, the back patches of the drop and 6.26 towards the upper patch.
Two resolutions of the mesh, 5 and 10 grid cells mm${}^{-1}$ at the finest part of the mesh, were used to validate the effect of the resolution.

The simulations were performed on a computer equipped with an Intel\textsuperscript{\textregistered} Core\textsuperscript{TM} i7-3960X CPU and with $32 \mathrm{GB}$ RAM.
The simulation results were rendered as movies using ParaView \cite{paraview}.
The interface between water and air was determined by thresholding the volume fraction of water at $\alpha$ = 0.5 to capture the center of transitional region between water and air.
Rendered movies were processed using the image processing pipeline that was also used to process the experimental results.

\section{Results and discussion}
\subsection{Assessment of the numerical result}
Here, the numerical results are assessed by comparison with the experimental results.

Before assessment of the results from a physical perspective, the effect of the mesh design was validated by evaluating the dependency on the mesh resolution.
No particular differences were observed in the results for the two different resolutions, which indicates that the mesh resolution has no significant effect on the result.
To inspect the numerical result with the fine resolution, results calculated with the finer mesh (10 cells / mm) were used for further analysis.

Figure\ \ref{fig:sequential images} shows sequential images of the experimental and simulated drops.
The experimental and simulated drops were comparable in that each drop exhibited a stable rebound.
The sequence of the deformation (spreading after first impact, forming a disk-shape, shrinking, making a head at the center of the disk, lift-off, shaking of the shape while in the air, and impacting again) was also reproduced in the calculation.
For high Weber numbers ($\Webe \geq 15$), the experimental result fluctuated, possibly due to asymmetrical expansion and contraction, while the numerical result was stable and had symmetrical expansion and contraction.
Deformation for the experimental drop was so sensitive that no symmetrical deformation could be achieved.
\begin{figure}[htbp]
    \begin{center}
        \includegraphics[width=3.25in]{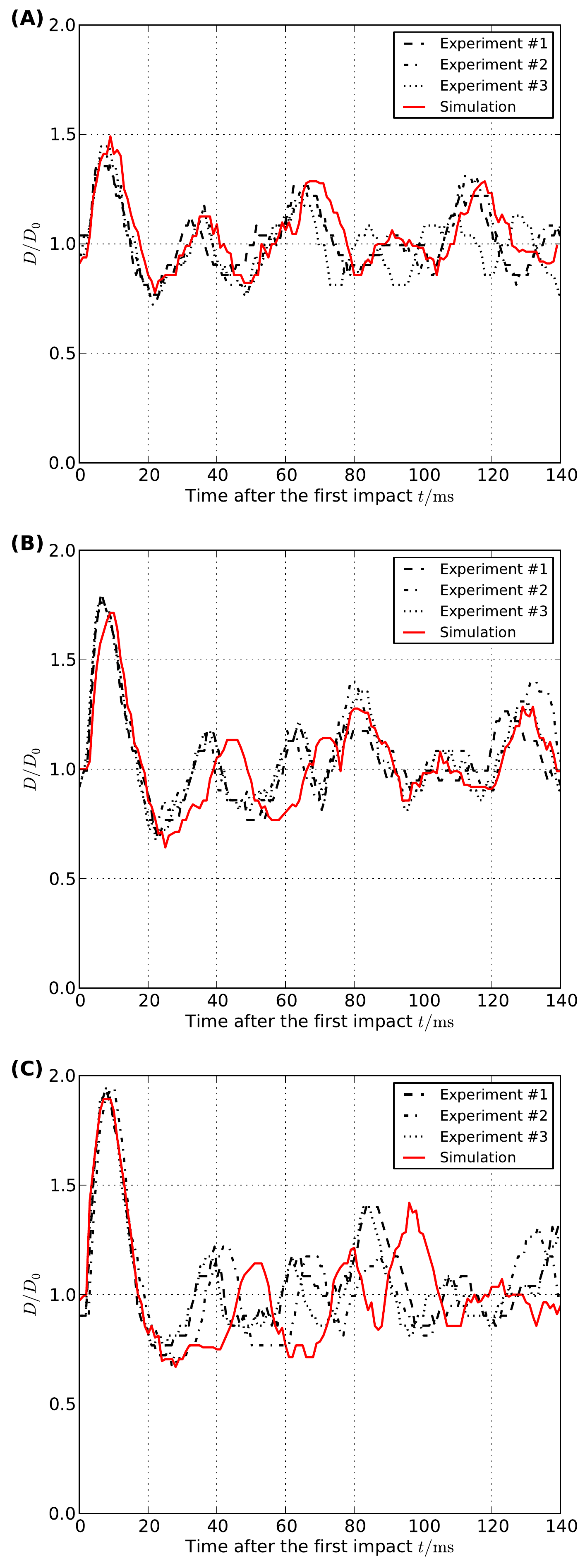}
        \caption{Time series of width for experimental and numerically simulated drops with (A) $\Webe=7$, (B) $\Webe=15$, and (C) $\Webe=23$. The first expansion and contraction (0-20 $\mathrm{ms}$) have good agreement.}
        \label{fig:WevsW_PWvsCFD}
    \end{center}
\end{figure}
\begin{figure}[htbp]
    \begin{center}
        \includegraphics[width=3.25in]{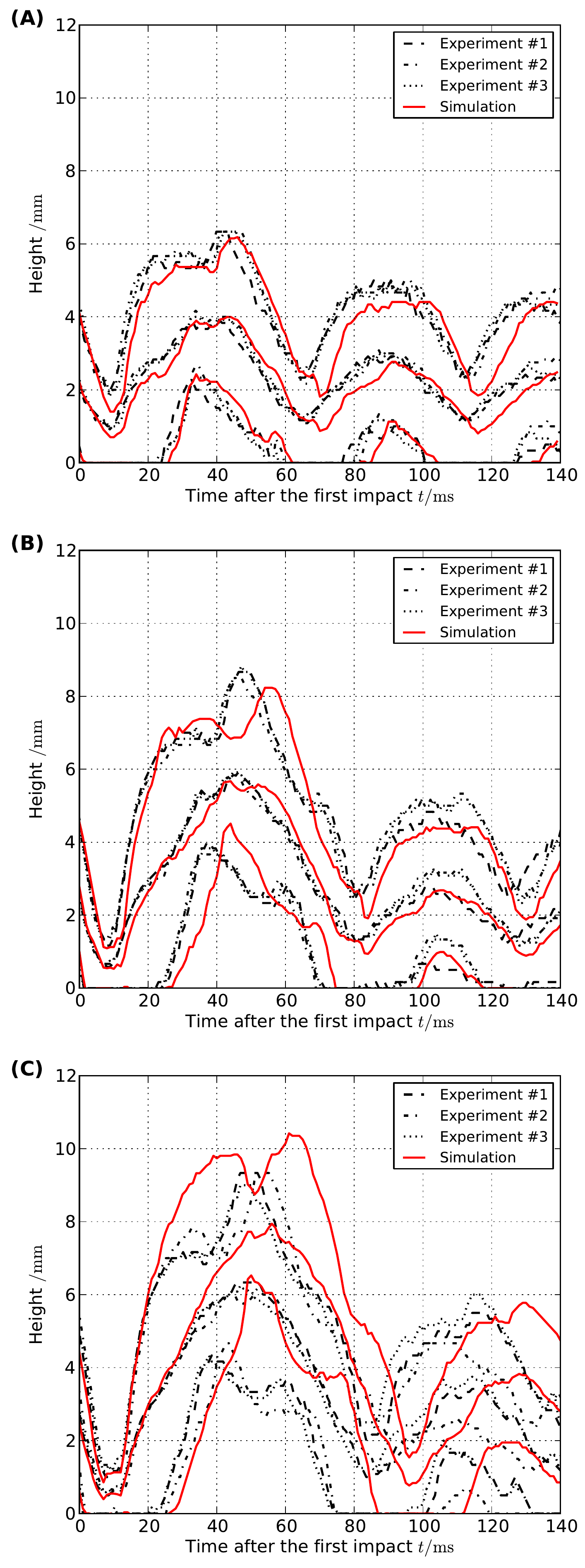}
        \caption{Time series of top height, middle height, and bottom height for experimental and numerically simulated drops with (A) $\Webe=7$, (B) $\Webe=15$, and (C) $\Webe=23$. The difference between the top and bottom heights represents deformation in the vertical direction and the middle height represents the approximate potential energy of the drop.}
        \label{fig:WevsH_PWvsCFD}
    \end{center}
\end{figure}
Time evolutions of the height and width of the drop are expected to provide vibrational patterns of the deformation process.
Time evolutions of relative diameter $D/D_0$ for both experimental and simulated drops with different Weber numbers were compared (Figure\ \ref{fig:WevsW_PWvsCFD}), and the top, middle, and bottom heights of the drops, $x_1$, $x_c$, and $x_2$, respectively (Figure \ref{fig:dropGeom} and Figure \ref{fig:WevsH_PWvsCFD}), show that both sets of results have the same vibrational patterns, although for high Weber numbers, the time spans between the first and second expansions and between the first and second impacts for the numerically simulated drops were slightly wider than those for the experimental drops.
The time series for the horizontal diameter of the drop during the impact approximately represents how much kinetic energy is converted to surface energy (Figure\ \ref{fig:WevsW_PWvsCFD}).
The time series for the middle height can be considered to represent approximately the potential energy of the drop.
Thus, as shown in Figure\ \ref{fig:WevsH_PWvsCFD}, the time series for the potential energy of the drop for both the experiments and the simulations can be considered to be in agreement.

The dissipated energy during the rebound is very difficult to determine because both the velocity and the surface area of the drop are unknown \cite{mundo1995}.
One effective way to experimentally estimate the dissipated mechanical energy is to calculate the ratio of the maximum height after the first impact to the initial height, as a ratio of mechanical energy at the maximum height to the initial mechanical energy:
\begin{equation}
    \frac{E_\mathrm{mech,hmax}}{E_\mathrm{mech,impact}} = \frac{H_\mathrm{max}}{H_0},
\end{equation}
by assuming that the potential energy is equal to the mechanical energy when the drop is at the highest position. 

\begin{figure}[h]
    \begin{center}
        \includegraphics[width=3.0in]{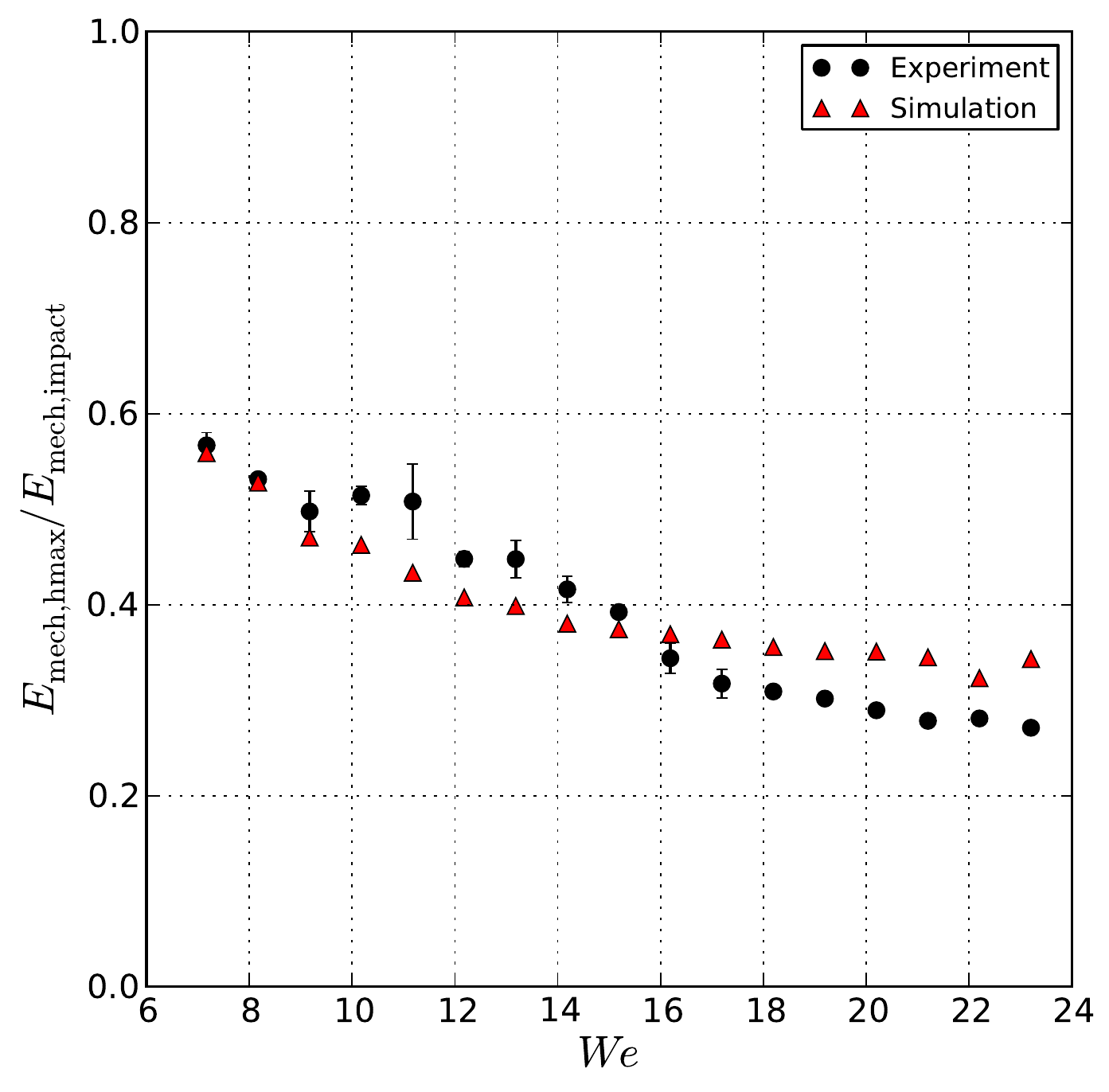}
        \caption{Weber number and the ratio of the mechanical energy of the drop at the maximum height after the impact to the initial mechanical energy. Both the numerical and experimental results showed a decrease in the ratio with an increase of the Weber number.}
        \label{fig:WevsEpsilon}
    \end{center}
\end{figure}

The ratio of mechanical energy at the maximum height to the initial mechanical energy for each Weber number is shown in Figure \ref{fig:WevsEpsilon}.
Both the numerical and experimental results showed a decrease with an increase of the Weber number.
The energy loss for the experimental result with high Weber numbers is considered to fluctuate due to asymmetrical deformation during the rebound (Figure \ref{fig:sequential images}).

Through the assessment performed here, the numerical result is considered to be reasonably reliable with respect to the deformation and dissipated energy.

\subsection{Quantitation of the energy conversion}
Kinetic energy and potential energy were calculated using the following respective equations:
\begin{align}
    E_\mathrm{kin} &= \int_\Omega \frac{1}{2} \rho U^{2} \mathrm{d}V, \\
    E_\mathrm{pot} &= \int_\Omega \rho gh \mathrm{d}V,
\end{align}
where $V$ is the volume and $\Omega$ is the entire domain for the calculation.
Under the condition that the width of the interface between water and air is asymptotically limited to zero, the integral over the interface can be reformulated by a volume with the gradient of the volume fraction, $\nabla \alpha$ \cite{brackbill1992}. Thus, the surface energy can be calculated using
\begin{equation}
    E_\mathrm{surf} = \int_\Omega \sigma |\nabla \alpha| \mathrm{d}V.
\end{equation}
The sum of the kinetic and potential energies is the mechanical energy:
\begin{equation}
    E_\mathrm{mech} = E_\mathrm{kin} + E_\mathrm{pot}.
\end{equation}
In this system, the pressure and volume are considered to be constant, and
the energy of interest is the sum of the mechanical and surface energies:
\begin{equation}
    E_\mathrm{ms} = E_\mathrm{mech} + E_\mathrm{surf}.
\end{equation}

\begin{figure}[htpb]
    \begin{center}
        \includegraphics[width=8.0cm]{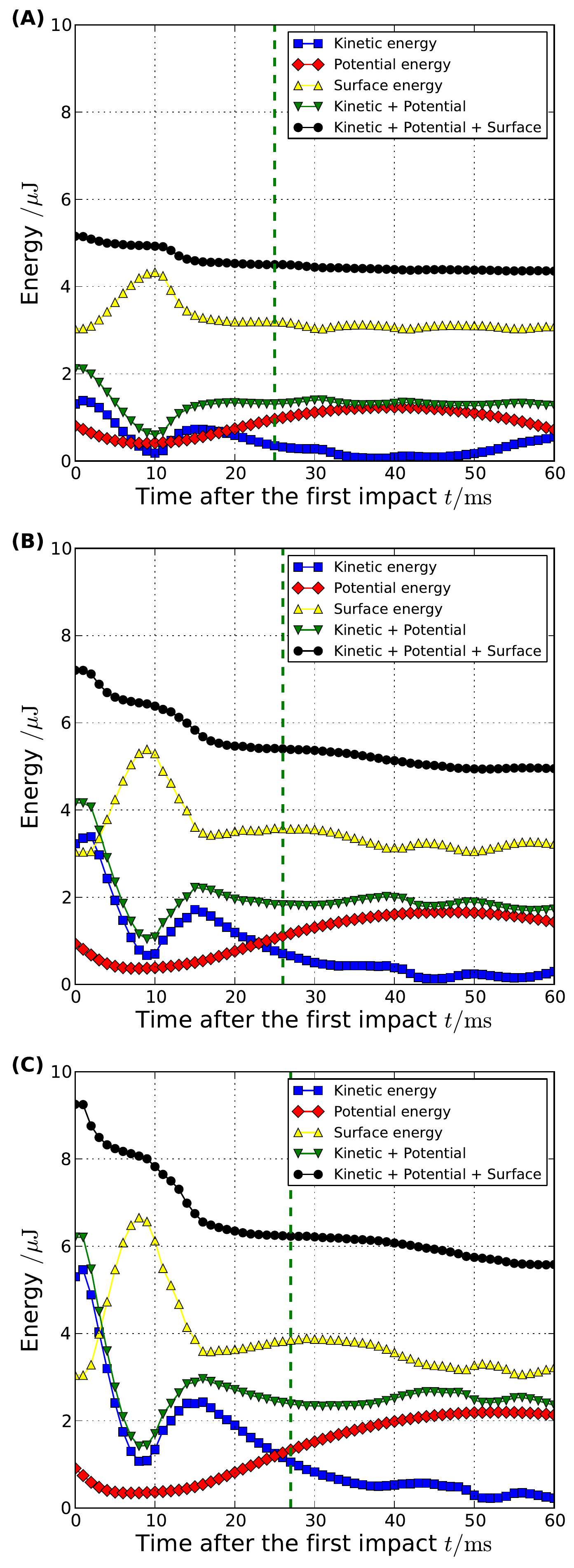}
        \caption{Kinetic, potential, and surface energies of drops as a function of $t$, the time after impact, with (A) $\Webe=7$, (B) $\Webe=15$, and (C) $\Webe=23$. The dashed line represents the transition point from regime I to regime II.}
        \label{fig:energiesWe07and23}
    \end{center}
\end{figure}

Figure\ \ref{fig:energiesWe07and23} shows the time evolution of the energies calculated from the numerical results.
At the impact ($t = 0\; \mathrm{ms}$), the mechanical energy begins to decrease rapidly and the surface energy simultaneously begins to increase.
When the surface energy reaches a maximum ($t \approx 8\; \mathrm{ms}$), the kinetic energy has a local minimum.
After reaching the maximum surface energy, the mechanical energy begins to increase while the surface energy decreases.
After takeoff of the drop ($t \approx 15\; \mathrm{ms}$), as evident for high Weber numbers, the conversion between the mechanical energy and surface energy still continues, which is considered to be caused by vibration of the drop in the air.
Interestingly, the changes of these energies cancel each other out and are considered to be conserved in the form of the sum of the mechanical and surface energies.

\subsection{Imaginary damped spring model}
A poorly elastic shock of a Leidenfrost drop has been modeled by an imaginary spring \cite{okumura2003,biance2006}, which is a linear spring model with two mass points that represent the mass of the drop at both ends of the spring.
Here, we extend the imaginary spring model by adding a damping term:
\begin{subnumcases}
    {}
        \frac{1}{2} m\frac{\mathrm{d}^2 x_1}{\mathrm{d}t^2} = - \frac{1}{2} mg - k\epsilon - c\frac{\mathrm{d}\epsilon}{\mathrm{d}t} & 
        \label{eq:imaginarySpring1} \\
        \frac{1}{2} m\frac{\mathrm{d}^2 x_2}{\mathrm{d}t^2} = - \frac{1}{2} mg + k\epsilon + c\frac{\mathrm{d}\epsilon}{\mathrm{d}t} + F, &
        \label{eq:imaginarySpring2}
\end{subnumcases}
where $x_1$ and $x_2$ are the heights of the bottom and top of the spring above the plate respectively, and
\begin{equation}
    \epsilon = x_1 - x_2 - D_0
    \label{eq:strainOfSpring}
\end{equation}
is the strain of the spring, $m$ is the mass of the drop, $D_0$ is the initial vertical length of the drop, $k$ is the stiffness of the spring, $c$ is the damping coefficient of the spring ($c \geq 0$), and $F$ is the external force loaded at the bottom of the spring.
Figure\ \ref{fig:dropGeom} shows a schematic diagram of the imaginary damped spring model.
Note that by combining Eqs. (\ref{eq:imaginarySpring1}) and (\ref{eq:imaginarySpring2}), the momentum equation for the centroid of the spring, $x_c = \frac{1}{2}(x_1 + x_2)$, can be represented as
\begin{equation}
	m\frac{\mathrm{d}^2x_c}{\mathrm{d}t^2} = -mg + F,
	\label{eq:momentumOfCentroid}
\end{equation}
which plots the free-fall and bounce-back of the spring.

Let us define the regime in which the drop is in contact with the vapor film over the plate as regime I.
In regime I, the height of the bottom of the spring is considered to be fixed ($x_2 = 0$); therefore,
\begin{equation}
	\frac{\mathrm{d}^2 x_1}{\mathrm{d}t^2} = \frac{\mathrm{d}^2 \epsilon_\mathrm{I}}{\mathrm{d}t^2},
\end{equation}
and
\begin{equation}
	\frac{\mathrm{d}^2 x_2}{\mathrm{d}t^2} = 0,
\end{equation}
where $\epsilon_\mathrm{I}$ is the strain in regime I.
Equations (\ref{eq:imaginarySpring1}) and (\ref{eq:imaginarySpring2}) then become:
\begin{subnumcases}
	{}
		\frac{\mathrm{d}^2 \epsilon_\mathrm{I}}{\mathrm{d}t^2} = - g - \frac{k_\mathrm{I}}{m}\epsilon_\mathrm{I} - \frac{c_\mathrm{I}}{m}\frac{\mathrm{d}\epsilon_\mathrm{I}}{\mathrm{d}t} &
		\label{eq:imaginarySpringI1} \\
		F = \frac{1}{2} \left( mg - k_\mathrm{I}\epsilon_\mathrm{I} - c_\mathrm{I}\frac{\mathrm{d}\epsilon_\mathrm{I}}{\mathrm{d}t} \right), &
		\label{eq:imaginarySpringI2}
\end{subnumcases}
where
\begin{equation}
	k_\mathrm{I} = 2k,\; c_\mathrm{I} = 2c.
\end{equation}
By solving Eq. (\ref{eq:imaginarySpringI1}), we obtain
\begin{equation}
	\epsilon_\mathrm{I} = -A_\mathrm{I} \mathrm{e}^{-\zeta_\mathrm{I} \omega_{\mathrm{I}} t_\mathrm{I}}\mathrm{sin}\left( \omega_{\mathrm{d,I}} t_{\mathrm{I}} + \psi_\mathrm{I}\right) - \frac{mg}{k_\mathrm{I}},
	\label{eq:epsilonI}
\end{equation}
where $t_\mathrm{I}$ is the time after the impact, $A_\mathrm{I}$ is the initial amplitude of the oscillation, $\zeta_\mathrm{I} = \frac{c_\mathrm{I}}{2\sqrt{mk_\mathrm{I}}}$ is the damping ratio, $\omega_{\mathrm{I}} = \sqrt{\frac{k_\mathrm{I}}{m}}$ is the undamped angular frequency of the spring, $\omega_\mathrm{d,I} = \sqrt{1 - \zeta_\mathrm{I}^2}\omega_\mathrm{I}$ is the under-damped harmonic oscillator, and $\psi_\mathrm{I}$ is the phase at the impact.

The time span from the lift-up to the next impact of the drop is defined as regime II.
In regime II, the bottom height of the spring is no longer fixed ($x_2 \geq 0$) and there is no external force loaded on the bottom mass point ($F = 0$).
Differentiation of Eq. (\ref{eq:strainOfSpring}) gives
\begin{equation}
	\frac{\mathrm{d}^2\epsilon_\mathrm{II}}{\mathrm{d}t^2} = \frac{\mathrm{d}^2 x_1}{\mathrm{d}t^2} - \frac{\mathrm{d}^2 x_2}{\mathrm{d}t^2},
	\label{eq:d2StrainII}
\end{equation}
where $\epsilon_\mathrm{II}$ is the strain in regime II.
The combination of Eqs. (\ref{eq:imaginarySpring1}), (\ref{eq:imaginarySpring2}), and (\ref{eq:d2StrainII}) gives
\begin{equation}
	\frac{\mathrm{d}^2\epsilon_\mathrm{II}}{\mathrm{d}t^2} = - \frac{k_\mathrm{II}}{m}\epsilon_\mathrm{II} - \frac{c_\mathrm{II}}{m}\frac{\mathrm{d}\epsilon_\mathrm{II}}{\mathrm{d}t},
	\label{eq:imaginarySpringII}
\end{equation}
where
\begin{equation}
	k_\mathrm{II} = 4k,\;c_\mathrm{II} = 4c.
\end{equation}
By solving Eq. (\ref{eq:imaginarySpringII}), we obtain
\begin{equation}
	\epsilon_\mathrm{II} = A_\mathrm{II} \mathrm{e}^{-\zeta_\mathrm{II} \omega_{\mathrm{II}} t_\mathrm{II}}\mathrm{cos}\left( \omega_{\mathrm{Id,I}} t_\mathrm{II} + \psi_\mathrm{II} \right),
\end{equation}
where $t_\mathrm{II}$ is the time after the lift-up, $A_\mathrm{II}$ is the amplitude of the oscillation, $\zeta_\mathrm{II} = \frac{c_\mathrm{II}}{2\sqrt{mk_\mathrm{II}}}$ is the damping ratio, $\omega_{\mathrm{II}} = \sqrt{\frac{k_\mathrm{II}}{m}}$ is the undamped angular frequency of the spring, $\omega_\mathrm{Id,I} = \sqrt{1 - \zeta_\mathrm{II}^2}\omega_\mathrm{II}$ is the under-damped harmonic oscillator, and $\psi_\mathrm{II}$ is the phase at lift-off.

The coefficients were obtained according to the description given in Appendix A.
The damping coefficient for regime I, $c_\mathrm{I}$, was determined to be $0.7\times10^{-3}\; \mathrm{kgs^{-1}}$ using Eq. (\ref{eq:viscousDampingCoeff}) with the result for $\Webe = 7$ and was reasonably assigned for all Weber numbers in this study, while that for regime II, $c_\mathrm{II}$, was determined to be half the value of $c_\mathrm{I}$. This difference of the damping coefficient indicates that the mechanism for energy loss is different between regimes I and II.
The stiffness $k$, determined by Eq. (\ref{eq:stiffness}), tends to decrease with an increase of the Weber number.
\begin{figure}[t]
	\begin{center}
		\includegraphics[width=8.6cm]{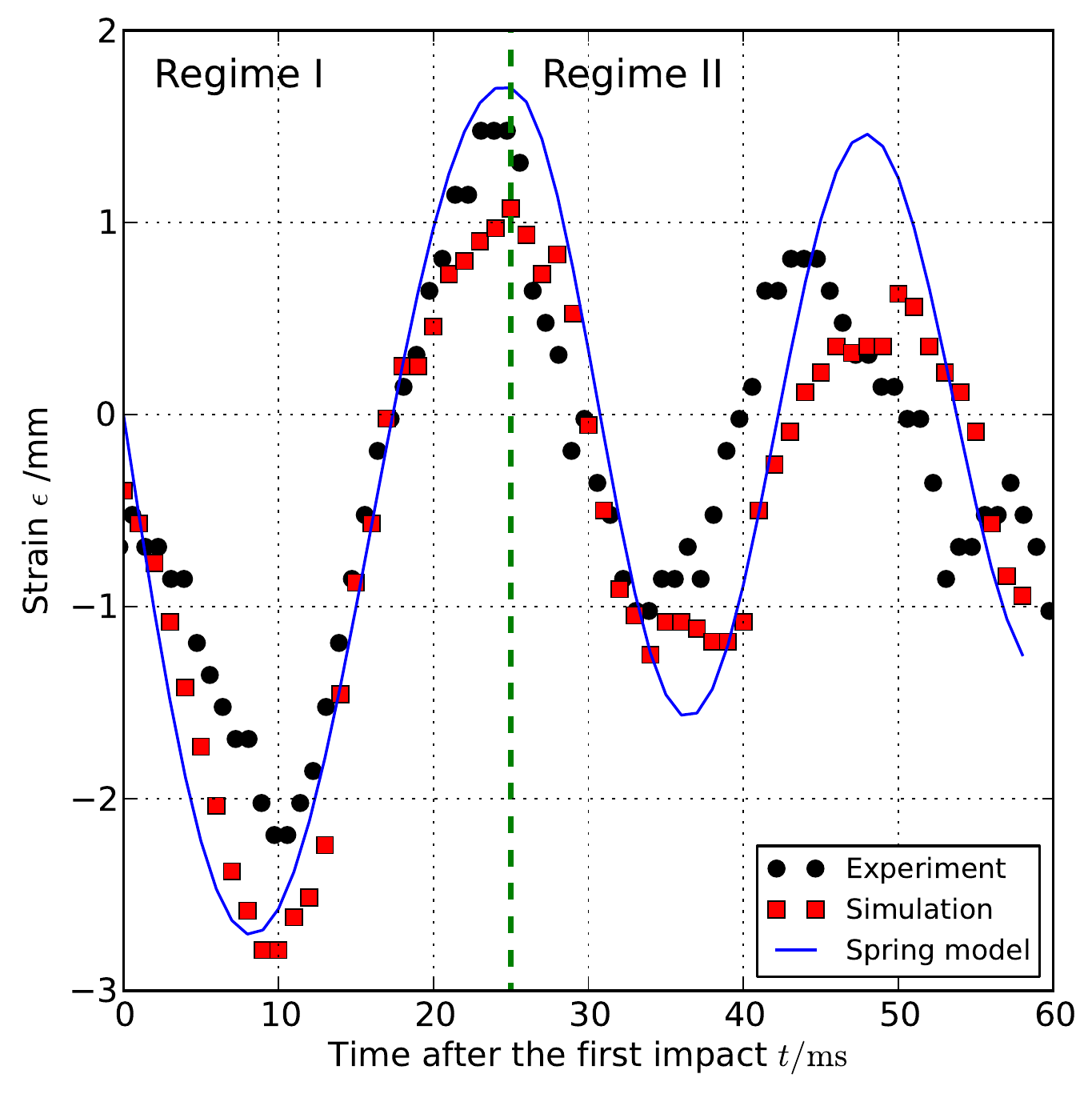}
		\caption{Time series of drop's vertical strain for the experiment, simulation, and the spring model with $\Webe = 7$.}
		\label{fig:timeVsVL}
	\end{center}
\end{figure}

\begin{figure}[htpb]
	\begin{center}
		\includegraphics[width=8.6cm]{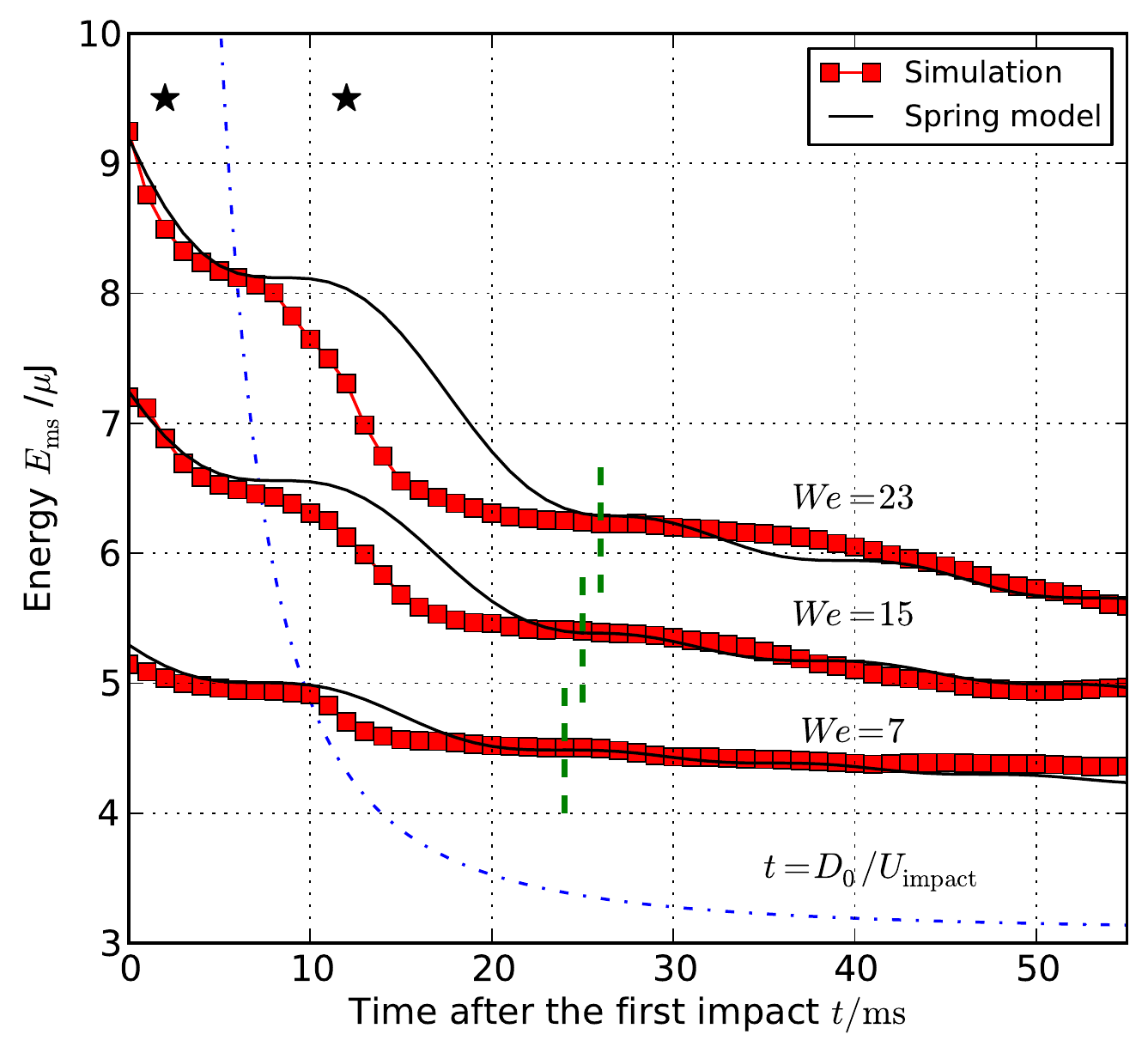}
		\caption{Time series for the sum of kinetic, potential, and surface energies from the simulation and spring model results. Each dashed line represents the transitional time from regime I to regime II. Major energy losses were observed at two moments, 2 ms and 12 ms after the impact as indicated by the "$\star$" marks. The dash-dotted line shows $t = D_0 / U_\mathrm{impact}$, which predicts the start time of the second energy decay of the drop.}
		\label{fig:timeVsEms}
	\end{center}
\end{figure}

The sum of the kinetic, potential, and elastic energies as the surface energy of the spring model can be calculated for each regime:
\begin{equation}
	\begin{split}
	E_\mathrm{ms,I} = &\frac{1}{4}m\left( \frac{\mathrm{d}\epsilon_\mathrm{I}}{\mathrm{d}t}\right)^2 + \frac{1}{2} mg(D_0 + \epsilon_\mathrm{I})\\
	& + \frac{1}{4}k_\mathrm{I}\epsilon_\mathrm{I}^2 + \sigma \pi D_0^2,
	\label{eq:energiesOfRegimeI}\\
	\end{split}
\end{equation}
\begin{equation}
	\begin{split}
	E_\mathrm{ms,II} = &\frac{1}{2}m\left( \frac{\mathrm{d}x_\mathrm{c,II}}{\mathrm{d}t}\right)^2 + \frac{1}{8}m\left( \frac{\mathrm{d}\epsilon_\mathrm{II}}{\mathrm{d}t_\mathrm{II}}\right)^2 + mgx_\mathrm{c,II}\\
	& + \frac{1}{8}k_\mathrm{II}\epsilon_\mathrm{II}^2 + \sigma \pi D_0^2,
	\label{eq:energiesOfRegimeII}
	\end{split}
\end{equation}
where $E_\mathrm{ms,I}$ and $E_\mathrm{ms,II}$ are sums of the mechanical energy and the surface energy in regime I and regime II, respectively.
These energies are shown in Figure\ \ref{fig:timeVsEms} with $\Webe = 7, 15, 23$.

\begin{figure}[htpb]
	\begin{center}
		\includegraphics[width=8.6cm]{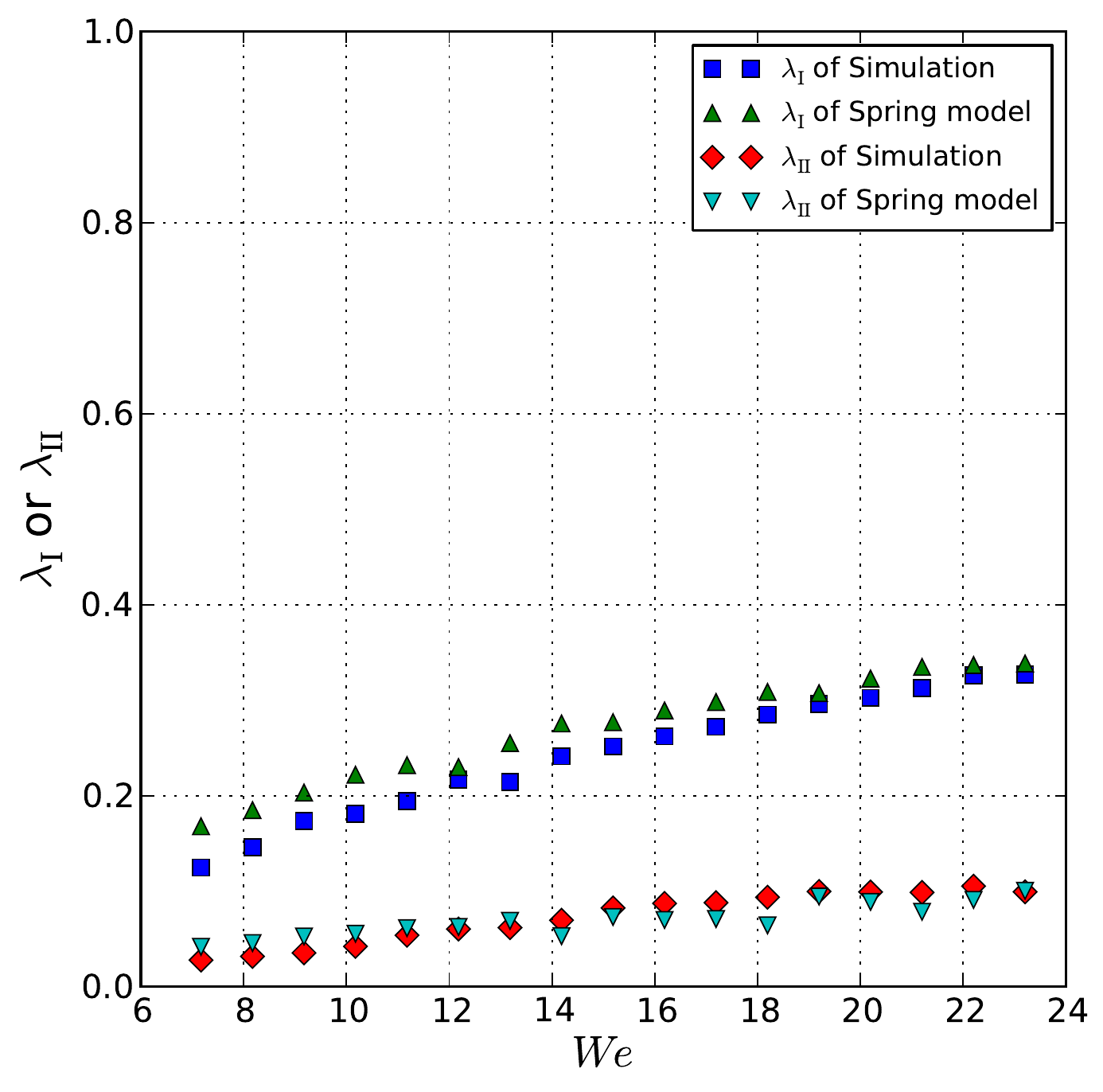}
		\caption{Energy loss rates over regime I $\lambda_\mathrm{I}$, and the energy loss rate over 1 cycle of oscillation in regime II $\lambda_\mathrm{II}$, with the simulated drop and the spring model.}
		\label{fig:timeVslambdaIandII}
	\end{center}
\end{figure}

The overall energy loss rate in regime I is expressed as
\begin{equation}
	\lambda_\mathrm{I} = 1 - \frac{E_\mathrm{ms,I}(t_\mathrm{II}=0)}{E_\mathrm{ms,I}(t_\mathrm{I}=0)},
\end{equation}
and the energy loss rate over 1 cycle of oscillation in regime II,
\begin{equation}
	\lambda_\mathrm{II} = 1 - \frac{E_\mathrm{ms,II}(t_\mathrm{II}=T_\mathrm{Id,I})}{E_\mathrm{ms,II}(t_\mathrm{II}=0)},
\end{equation}
corresponds well for both the simulation and the spring model (Figure\ \ref{fig:timeVslambdaIandII}).

While the vertical strain and energy loss rates of the drop were well explained by the spring model (Figures \ref{fig:timeVsVL} and \ref{fig:timeVslambdaIandII}), the second decrease of the spring model lagged that of the simulated drop (Figure \ref{fig:timeVsEms}). This time lag indicates that true damping factor has an other period than the damping term of the spring model.

\begin{figure*}[htbp]
	\begin{center}
		\includegraphics[width=\textwidth]{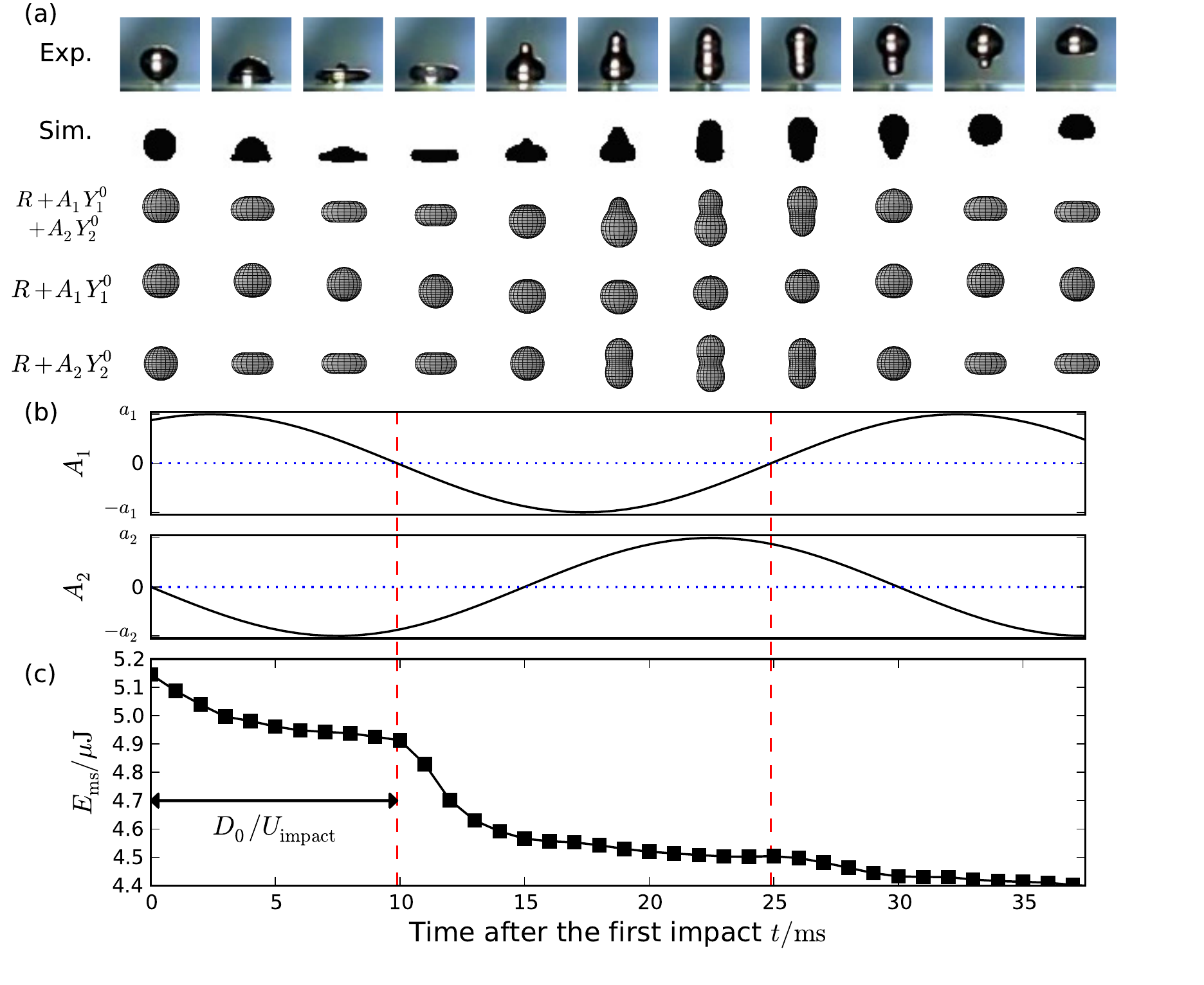}
		\caption{A sequential analysis of the deformation of the impacting drop with shared time axis, with $\Webe = 7$. (a) Spherical harmonic deformations represented by $R + A_1 Y_1^0 + A_2 Y_2^0$, $R + A_1 Y_1^0$, and $R + A_2 Y_2^0$ with $R = 1.85$ mm, $a_1 = 2$ mm, and $a_2 = 2$ mm, together with that of drops of the experiment (Exp.) and the CFD simulation (Sim.). (b) Intensities of $Y_1^0$ and $Y_2^0$, $A_1$ and $A_2$, respectively, are plotted as functions of the time after the first impact, $t$. (c) Sum of mechanical and surface tension of simulated drop is plotted as a funcion of $t$. Dashed lines in (b) indicates moments of $A_1 = 0$, at which exponentially energy decays starts as shown in (c).} 
		\label{fig:timeline}
	\end{center}
\end{figure*}

\begin{figure}[htpb]
	\begin{center}
        \includegraphics[width=7.8cm]{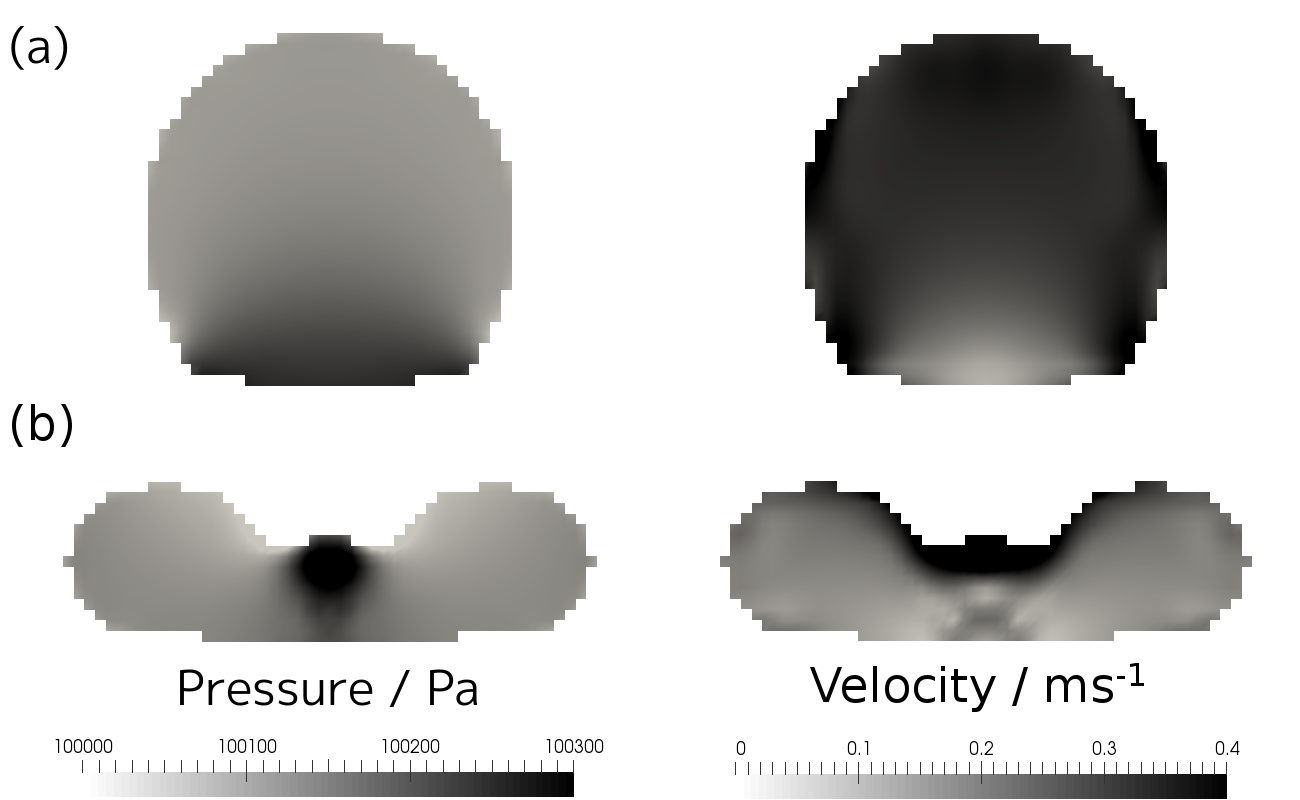}
		\label{fig:againstPressureForce}
		\caption{Pressure (left) and magnitude of the velocity (right) inside the drop, from a cross-sectional lateral view at the drop center ($\Webe = 7$). (a) At the beginning of the impact ($t = 2$ ms), a pressure surge occurs at the bottom part of the drop, where magnitude of the velocity is nearly zero. and (b) At the retraction from the disk-shaped drop ($t = 12$ ms), the pressure-induced motion suppresses the retracting motion.}
		\label{fig:againstPressureForce}
	\end{center}
\end{figure}

\begin{figure}[htpb]
	\begin{center}
		\includegraphics[width=8.6cm]{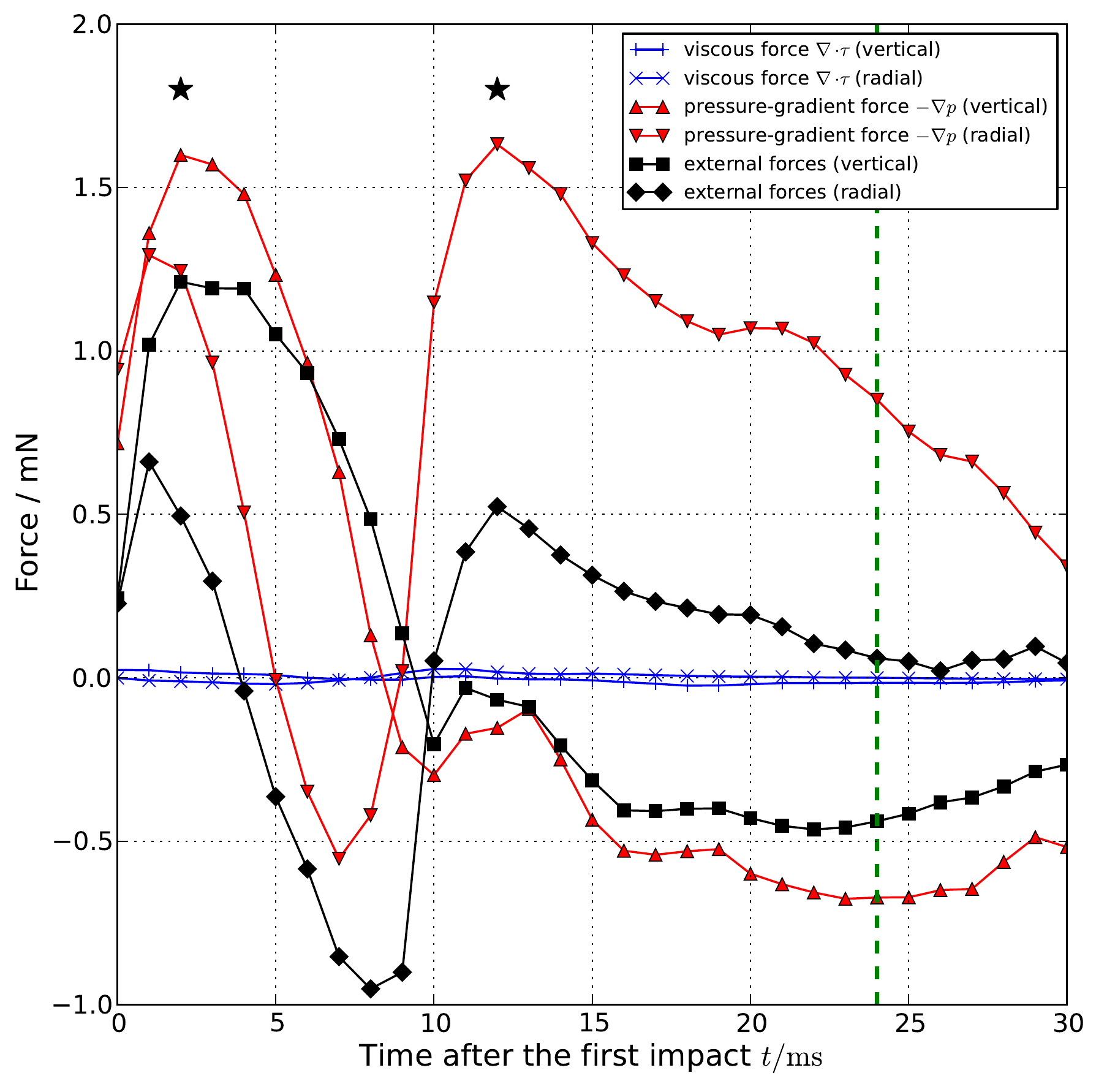}
        \caption{The viscous force $\nabla \cdot \boldsymbol{\tau}$, pressure-gradient force $-\nabla p$, and the external forces (right-hand side of Eq. (\ref{eq:momentum equation})) damping the motion of the drop for the vertical and radial directions were obtained from the simulated results ($\Webe = 7$). The dashed line represents the transition point from regime I to regime II. The pressure-gradient force dominates the external force and the contribution of the viscous force is fairly small. At the two major energy losses ($\star$), the external forces damp the velocity, the former against the vertical direction and the latter against the radial direction.}
		\label{fig:timeVsForces}
	\end{center}
\end{figure}

\subsection{Spherical harmonic analysis}
A water drop on an oscillating plate \cite{okada2006, takaki1984, takaki1985} or on a superheated plate \cite{yoshiyasu1993} is known to show spherical harmonic oscillation due to the surface tension force.
We will compare the deformations of the drop on the impact with the spherical harmonic oscillation.

An oscillating drop can be represented as a linear combination of spherical harmonic functions as \cite{lamb1945,prosperetti1980,landau1987}
\begin{equation}
	r(t, \theta, \phi) = R + \sum_{n,m} A_n (t) Y_{n}^{m} (\theta, \phi),
	\label{eq:spherical_harmonic_oscillation}
\end{equation}
where $R$ is the radius of an unperturbed sphere-shape drop, $A_n (t) = a_n \mathrm{cos} (\omega_n t + \psi_n)$ are intensities of the modes, and $Y_{n}^{m}$ are Laplace's spherical harmonics with orders $n = 0,1,2,\cdots$, and degrees $m = -n, -(n-1), \cdots, n-1, n$. 

The deformation of the drop in this study is radially symmetric, thus we only consider Laplace's spherical harmonics with 0 degree, $Y_{n}^{0}$.
Furthermore, we assume that the deformation of the impacting drop is aligned in vertical or radial axes, so oscillation modes can be narrowed down to $n = 1$ and $n = 2$.
Therefore Eq. (\ref{eq:spherical_harmonic_oscillation}) is simplified as
\begin{equation}
	r(t, \theta, \phi) = R + A_1 (t) Y_{1}^{0} (\theta, \phi) + A_2 (t) Y_{2}^{0} (\theta, \phi).
	\label{eq:spherical_harmonic_oscillation_n12}
\end{equation}

Both of $Y_{1}^{0}$ and $Y_{2}^{0}$ modes are radially symmetric, but $Y_{2}^{0}$ is horizontally symmetric while $Y_{1}^{0}$ is horizontally antisymmetric, as shown in sequential images of $R+A_1 Y_1^0$ and $R+A_2 Y_2^0$ in Figure \ref{fig:timeline}(a).
Using this difference, harmonic phase shifts of these two modes, $\psi_1$ and $\psi_2$ are determined by discriminating the symmetry of deformations of the drop.

Immediately after the impact ($t = 0$ ms), the drop is in a spherical shape, thus the phase of mode $n = 2$ must be $\frac{\pi}{2}$ or $-\frac{\pi}{2}$.
After impacting on the plate, the drop spreads radially and forms into a disk-shape, which corresponds to $A_{2} < 0$, therefore $\psi_2 = \frac{\pi}{2}$.

The determination of the harmonic phase of the $Y_1$ mode is more difficult than the $Y_2$ mode.
At the moment of the maximum width of spreading drop, the drop is in a disk-shape which is horizontally symmetric, therefore $A_1 = 0$.
From this moment to the lift-off, the drop forms an antisymmetric matryoshka-shape, which indicates $A_1 < 0$.
After the lift-off, the drop forms into a vertical peanut-shape (e.g. 25 ms) which is horizontally symmetric thus $A_1 = 0$.
By extrapolating from these conditions, $A_1$ must be a positive value at $t = 0$.
By looking carefully at energy decay curves in Figure \ref{fig:timeVsEms}, we found that the second energy decay starts slightly earlier at increasing Weber number.
Let us consider a free end reflection of the drop as a pulse over the vertical axis on the bottom plate as a free end.
We envisage that after the impact the drop receives the reflection of its impact velocity and induces a vertical uplift (i.e. $A_1 > 0$), until the pulse passes over the end.
Therefore, a length of the time during which the drop passes over its diameter with the impact velocity, $D_0/U_\mathrm{impact}$, is considered to characterize the phase of the $Y^0_1$ mode.
As shown by the dash-dotted line in Figure \ref{fig:timeVsEms}, $t = D_0/U_\mathrm{impact}$ predicts beginning time of the second decay well.

The deformation of Eq. (\ref{eq:spherical_harmonic_oscillation_n12}) with determined phases $\psi_1$ and $\psi_2$ together with experimental and simulated drops is shown in Figure \ref{fig:timeline}(a).
The deformation sequence of spherical shape, disk-shape, matrioshka-shape, and peanut-shape, is reproduced by Eq. (\ref{eq:spherical_harmonic_oscillation_n12}).

$A_1$ and $A_2$ as functions of time after the impact are shown in Figure \ref{fig:timeline}(b).
Actually, the imaginary spring model represents the $Y_2^0$ mode: $A_2$ has same vibration mode with the drop's vertical strains, $\epsilon$, which is shown in Figure \ref{fig:timeVsVL}. This is because the vertical strain of a $Y_{1}^{0}$ mode is always zero and thus height of the shape represented by Eq. (\ref{eq:spherical_harmonic_oscillation_n12}) only depends on the $Y_2^0$ mode.

One of the most important insights from the spherical harmonic analysis is the existence of an another vibrational mode, $Y_1^0$ in the deformation of the impacting drop, which is not considered in the spring model.

\subsection{Energy loss upon the impact}
Time series of $A_1$ and $A_2$ were compared with that of the total energy $E_\mathrm{ms}$ of the simulated drop with $\Webe = 7$ (Figures \ref{fig:timeline}(b) and \ref{fig:timeline}(c)).
We found that the cycle of the $Y^0_1$ mode is synchronous with that of the repetitive energy decay.
At the time spans when $Y_1^0$ increases its amplitude ($\left| \frac{\mathrm{d}A_1}{\mathrm{d}t}\right| > 0$), the energy starts an exponential decay.
Figure\ \ref{fig:timeVsEms} shows that the damped imaginary spring model predicts lagged second decay behind the simulated drop.
As mentioned in section 4.4, the spring model represents the $Y_2^0$ mode.
Considering that the cycle of the energy decay depends on the  $Y_1^0$ mode and the spring model does not consider the $Y_1^0$ mode, the mismatch of the decay timing can be explained.

When the drop impacts at the bottom plate, the bottom part of the drop is forced to stop and the vertical velocity is forced to be zero suddenly.
This sudden change of the velocity induces the pressure surge (i.e., stagnation pressure).
Figure \ref{fig:againstPressureForce}(a) shows that a pressure surge occurs at the beginning of the impact.
The amount of the pressure surge is roughly estimated as $\rho D_0 U_\mathrm{impact}/\Delta t \sim 200$ Pa, where $\Delta t \approx D_0/U_\mathrm{impact} \sim 10$ ms is the time span to stop the free-fall motion of the drop.
This phenomenon is similar to a water hammer with slow valve closure (slower than sound propagation), in which a pressure surge occurs when a fluid in motion is forced to stop.
The pressure surge accompanies a pressure-gradient force, which is expressed by the term $-\nabla p$ in the Eq. (\ref{eq:momentum equation}).

In many cases of fluid dynamics, a pressure-gradient force is a driving force of a flow (e.g., a channel flow).
However, in this case, the pressure-gradient force consequently dumps the motion inside the drop. The pressure-induced force associated with the stagnation pressure is an important factor in the Drop Deformation and Breakup (DDB) model \cite{ibrahim1993,vargas2012} introduced by Ibrahim et al., which successfully predicts the deformation of spray drops.

The pressure-gradient force generated at the impact induces an upward flow inside the drop in the time span of $D_0/U_\mathrm{impact}$ from the impact. The upward flow is subsequently reflected at the end of the drop due to the surface tension.
Therefore, the flow induced by the pressure-gradient force generates a vertical vibrational motion.
The $Y_1^0$ mode found in our spherical harmonics analysis represents this vertical vibrational motion.

Meanwhile after the impact, a major part of the free-fall motion of the drop is converted to radially spreading motion, which is also subsequently reflected at the end of the drop due to the surface tension.
This motion forms a vibrational motion represented by the $Y_2^0$ mode.
Therefore, the free-fall motion of the drop before the impact is converted to two motions after the impact, the pressure-induced motion and inertial motion, represented by $Y_1^0$ and $Y_2^0$ modes, respectively.

To investigate the breakdown of the force acting to dump inside the drop, viscous force, pressure-gradient force, and total external force (viscous force, pressure-gradient force, surface tension force, and gravitational force) were summed inside the drop of the CFD simulation for each time step.
Specifically, we calculated an $\alpha$-weighted summation of each force over grid cells with a negative value of inner product of a force and velocity, as expressed in $\sum_{i, \mathbf{f}_i \cdot \mathbf{U}_i < 0} \alpha_i \mathbf{f}_i$,
where $i$ is an index of grid cells, $\mathbf{f}_i$ is a force at $i$-th cell, $\mathbf{U}_i$ is velocity at $i$-th cell, and $\alpha_i$ is the volume fraction of water phase at $i$-th cell.
Considering the deformation of the bounce of the 3D drop where spreading and shrinking in vertical and radial directions, these forces were split into vertical and radial components. The radial force component shows major forces within the drop in a disk-shape, while the vertical component shows ones within the drop in a cylinder-shape.
Figure\ \ref{fig:timeVsForces} shows that amongst the forces damping the motion of the drop, the pressure-gradient force dominates the external forces (right-hand side of Eq. (\ref{eq:momentum equation})) and the viscosity effect is fairly small.
The small impact of the viscosity on the drop deformation was also reported by Renardy {\it et al.} \cite{renardy2003}.

The pressure-induced motion resists the free-fall inertial motion at the beginning of the impact ($t = 2$ ms), and then resists the inertial motion when the drop is retracting from the disk-shape ($t = 12$ ms).
Figure \ref{fig:timeVsForces} shows that the pressure-gradient force dominantly resists the motion of the drop at these two timings.
At the retraction of the drop, the direction of the pressure-induced motion is inherently downward, however, due to the disk-shaped drop as a flow field and the existence of the bottom plate, it is forced to advance radially (Figure \ref{fig:againstPressureForce}(b)).
As the result, the retracting inertial motion of the drop is dumped by this radially spreading pressure-induced motion.

Most part of the pressure-induced motion decays during the two resistances to the inertial motion, however, still remains with a small intensity after the lift-off, causing a small energy loss starting at $t = 25$ ms as shown in Figures \ref{fig:timeVsEms} and \ref{fig:timeline}(c).

Note that we have considered just the first dry-rebound.
Biance \textit{et al.} \cite{biance2006} have shown restitution coefficients of successive dry-rebounds of a drop with diameter of 1 mm.
They reported that the restitution coefficient $e$ is relatively low at the first impact ($e \sim \Webe^{-1/2}$, called as poorly elastic shocks) but very close to 1 after multiple bounces (called as quasi-elastic shocks).
With lower Weber number, the energy loss tends to be small as shown in Figure \ref{fig:WevsEpsilon}.
Thus, one reason of the small energy loss after multiple bounce is the low Weber number.
Biance \textit{et al.} also reported that in the quasi-elastic shocks the vibration of drop's diameter is in phase with the flight of the drop.
The other reason of the small energy loss after multiple bounce is considered that the pressure surge disappears due to the inertial motion synchronized with the bounce of the drop.
The synchronized $Y^0_2$ mode, which suppresses the impact velocity, avoids the sudden stop at the bottom and generating the pressure surge.

\section{Conclusion}
The dynamics of a dry-rebounding drop was quantitatively obtained from numerically simulated results that were assessed with respect to experimental results.
The dynamics was quantitatively explained with an imaginary damped spring model, however, the second energy decay predicted by the spring model was lagged behind the simulated drop, which indicates that the true damping factor is other than the damping term of the spring model.

From the analysis of the spherical harmonic deformation, we found that the deformation is a combination of $Y_1^0$ and $Y_2^0$ modes.
The cycle of the $Y_1^0$ mode was synchronous with that of the energy decay, which indicates that the decay timing depends on the $Y_1^0$ mode, rather than the $Y_2^0$ mode represented by the spring model.

At the beginning of the impact, the bottom part of the drop is forced to stop suddenly, which induces a pressure surge.
From the simulated results, the pressure surge was actually found.
The pressure-gradient force associated with the pressure surge induces a upward motion.
At the same time, the free-fall motion is converted to the radially spreading motion.
These two motions form two different vibrational modes due to the surface tension.
The pressure-induced motion and inertial motion correspond to the $Y_1^0$ and $Y_2^0$ modes, respectively.

Analysis of the forces damping the motion of the drop suggested that the viscous impact on the drop is fairly small.
Considering that the $Y_1^0$ mode was synchronous with the repetitive exponential decay of the energy and the pressure-gradient force dominantly resists the motion of the drop, we conclude that the pressure-induced motion dumps the inertial motion.

\section{Acknowledgements}
A. Nishimura would like to thank Prof. Tsuguki Kinoshita for suggestions on the energy conversion.
A. Nishimura would like to thank Dr. Chris Greenshields for training him in how to use OpenFOAM.

\appendix
\section{Coefficients of the imaginary damped spring model}
To conserve the kinetic energy at impact, the impact speed of the spring, $U_\mathrm{impact}^*$, is recalculated from the impact speed of the drop, $U_\mathrm{impact}$, and applied for mass point 1 because mass point 2 cannot move:
\begin{equation}
	U_\mathrm{impact}^* = \sqrt{2} U_\mathrm{impact}.
\end{equation}

Assuming that the strain of the spring has a maximal value at lift-off, the under-damped angular frequency, $\omega_\mathrm{d,I}$, can be obtained by the period, $T_\mathrm{d,I}$, which is equal to twice the length of time from the minimum $\epsilon$ to the maximum $\epsilon$:
\begin{equation}
	\omega_\mathrm{d,I} = \frac{2\pi}{T_\mathrm{d,I}}.
\end{equation}

The time derivative of the strain at impact is equal to the impact speed:
\begin{equation}
	\left. \frac{\mathrm{d}\epsilon_\mathrm{I}}{\mathrm{d}t}\right|_{t_\mathrm{I}=0} = - U_\mathrm{impact}^*,
\end{equation}
which leads, by assuming $\mathrm{sin}(\psi_\mathrm{I}) \approx 0$ and $\mathrm{cos}(\psi_\mathrm{I}) \approx 1$, to the amplitude
\begin{equation}
	A_\mathrm{I} = \frac{U_\mathrm{impact}^*}{\omega_\mathrm{d,I}}.
\end{equation}

At the first moment of impact, assuming that mass point 1 continues to move at $U_\mathrm{impact}^*$, then $\frac{\mathrm{d}^2 \epsilon_\mathrm{I}}{\mathrm{d} t^2} = 0$, and then from Eq. (\ref{eq:imaginarySpringI1}), we obtain
\begin{equation}
	c_\mathrm{I} = \frac{mg}{U_\mathrm{impact}^*}.
	\label{eq:viscousDampingCoeff}
\end{equation}

The stiffness, $k_\mathrm{I}$, can be obtained from
\begin{equation}
	k_\mathrm{I} = \frac{1}{2} \left( m\omega_\mathrm{d,I}^2 + \sqrt{m^2 \omega_\mathrm{d,I}^4 + \frac{c_\mathrm{I}^4}{4m^2}}\right).
	\label{eq:stiffness}
\end{equation}

At the start of contact of the bottom mass point of the spring ($t_\mathrm{I} = 0$), the strain is considered to be zero ($\epsilon_\mathrm{I} = 0$); therefore, the value of $\psi_\mathrm{I}$ can be obtained:
\begin{equation}
	\psi_\mathrm{I} = \mathrm{arcsin}\left( -\frac{mg}{k_\mathrm{I}A_\mathrm{I}}\right).
\end{equation}
Assuming that the strain reaches a maximum at lift-off ($t_\mathrm{II} = 0$),
\begin{equation}
	\frac{\mathrm{d}\epsilon_\mathrm{II}}{\mathrm{d}t} = 0,
\end{equation}
we obtain
\begin{equation}
\psi_\mathrm{II} = 0.
\end{equation}

The centroid of the spring in regime II, $x_\mathrm{c,II}$, is
\begin{equation}
	x_\mathrm{c,II} = U_\mathrm{lift\mathchar`-off}t_{\mathrm{II}} - \frac{1}{2}gt_{\mathrm{II}}^2 + x_\mathrm{c,lift\mathchar`-off},
\end{equation}
and the velocity of the centroid of the spring is
\begin{equation}
	\frac{\mathrm{d}x_\mathrm{c,II}}{\mathrm{d}t} = U_\mathrm{lift\mathchar`-off}t_\mathrm{II} - gt_\mathrm{II},
\end{equation}
where $U_\mathrm{lift\mathchar`-off}$ and $x_\mathrm{c,lift\mathchar`-off}$ are respectively the velocity and the position of the centroid of the spring at lift-off.
$U_\mathrm{lift\mathchar`-off}$ was obtained from
\begin{equation}
	U_\mathrm{lift\mathchar`-off} = \frac{1}{\sqrt{2}}\left.\frac{\mathrm{d}\epsilon_\mathrm{I}}{\mathrm{d}t}\right|_{t_\mathrm{II}=0}.
\end{equation}

\bibliographystyle{apacite-noand} 

\end{document}